\documentclass[preprint,
amsmath,amssymb,
 aps, prd 
]{revtex4-1}

\usepackage{graphicx}
\usepackage{dcolumn}
\usepackage{bm}
\usepackage{hyperref}

\DeclareMathOperator{\arcsinh}{arcsinh}

\begin{document}


\title{Anomalous and Linear Holographic Hard Wall Models for \\ Glueballs and the Pomeron}

\author{Rafael A. Costa-Silva}
 \email{rafaelcosta@pos.if.ufrj.br}
 \email{rc-fis@outlook.com}
\author{Henrique Boschi-Filho}%
 \email{boschi@if.ufrj.br}
 \email{hboschi@gmail.com}
\affiliation{Instituto de Física, Universidade Federal do Rio de Janeiro, 21.941-909 - Rio de Janeiro - RJ - Brazil
}%

\date{\today}

\begin{abstract}
In this work we propose improved holographic hard wall (HW) models by the inclusion of anomalous dimensions in the dual operators that describe glueballs inspired by the AdS/CFT correspondence. The anomalous dimensions come from well known semi-classical gauge/string duality analysis showing a dependence with the logarithm of spin $S$ of the boundary states. We show that these logarithm anomalous dimensions of the high spin operators combined with the usual HW model allow us to match the pomeron trajectory and give glueball masses which are better than that of the original HW and soft wall (SW) models in comparison with lattice data. We also build up other anomalous HW  (AHW) models considering that the logarithm anomalous dimensions can be approximated by a truncated series of odd powers of the difference $\sqrt{S}-1/\sqrt{S}$. These models also fit the pomeron trajectory and produce good glueball masses. Then, we consider an anomalous dimension which is proportional to $\sqrt{S}$, providing reasonable results. Finally, we propose an asymptotic linear AHW model which effective dimensions for high spins operators are of the form $\Delta=a\sqrt{S}+b$, where $a$ and $b$ are constants to be fixed by comparison with the soft pomeron trajectory. In this last model, the Regge trajectory is asymptotically linear even for very high spins ($J\sim 100$) matching the soft pomeron trajectory accurately and generates glueball masses with deviations with respect to the lattice data better than the original HW and SW models. 
\end{abstract}

\maketitle


\tableofcontents

\section{Introduction}

QCD describes strong interactions. At high energies its coupling is small so that it can be treated perturbatively. On the other hand, at low energies, the QCD coupling is large and non-perturbative methods are needed to tackle phenomena like confinement, phase transitions, and hadronic spectra. This non-perturbative behavior usually requires involved numerical calculations known as lattice QCD. Alternatively, low energy QCD may be approached by other methods, as the solution of Schwinger-Dyson equations, QCD sum rules and effective models (for a review see, {\it e.g.},  \cite{Gross:2022hyw}). 
In particular, models inspired by the AdS/CFT correspondence \cite{Aharony:1999ti, Polchinski:2001tt, Polchinski:2002jw, Gubser:2002tv, Ramallo:2013bua} 
proved useful to describe different aspects of hadrons with various spins, as  glueballs  \cite{Csaki:1998qr, deMelloKoch:1998vqw, Hashimoto:1998if, Csaki:1998cb, Minahan:1998tm, Brower:2000rp, Caceres:2000qe, Boschi-Filho:2002xih, Boschi-Filho:2002wdj, Apreda:2003sy, Amador:2004pz, Evans:2005ip, Caceres:2005yx, Boschi-Filho:2005xct, Colangelo:2007pt,  FolcoCapossoli:2013eao, FolcoCapossoli:2015jnm, FolcoCapossoli:2016uns, Capossoli:2021ope}, 
as well as for mesons and baryons, as for instance, in 
\cite{Sakai:2004cn, Sakai:2005yt, 
deTeramond:2005su, Ghoroku:2005vt, 
 Erlich:2005qh, DaRold:2005mxj, Karch:2006pv, Brodsky:2006uqa, 
 Hata:2007mb, Forkel:2007cm, 
 Gursoy:2007cb, Gursoy:2007er, 
 Erdmenger:2007cm, Colangelo:2008us, dePaula:2008fp, Vega:2008af, Abidin:2009hr, Gutsche:2011vb, Li:2012ay, 
 Brodsky:2014yha, 
 Sonnenschein:2016pim, 
 FolcoCapossoli:2019imm, Afonin:2020msa, 
Rinaldi:2021dxh}.

The holographic HW   model \cite{Boschi-Filho:2002xih, Boschi-Filho:2002wdj} introduces a hard cutoff in the AdS space, this way, hadronic masses $M$ are proportional to the zeros of  Bessel functions, $J_\nu(z)$, {\sl i.e.}, proportional to the value of the argument $z$ where the corresponding Bessel function is zero. This model was inspired by holographic  descriptions of hard scattering of glueballs \cite{Polchinski:2001tt} and deep inelastic scattering of hadrons \cite{Polchinski:2002jw}, and it is very useful to obtain hadronic form factors, structure functions, parton distribution functions, etc (see, {\sl e.g.}, 
\cite{Grigoryan:2007vg, Mamo:2021cle}).   It is important to mention that in the HW, the order of the Bessel function $\nu$ is related to the conformal dimension of the dual operator. For instance, for even spin $S$ glueballs, it reads $\nu= S + 2 $ \cite{deTeramond:2005su, Boschi-Filho:2005xct}.
Actually, in Ref. \cite{Boschi-Filho:2005xct},  approximate linear Regge trajectories, $J\times M^2$, for light even glueballs were obtained $\alpha(t=M^2)=(0.80 \pm 0.40) + (0.26\pm0.02)M^2$ and compared to a reasonable  approximation to that of the soft pomeron $\alpha(t=M^2)=1.08 + 0.25 M^2$ \cite{Landshoff:2001pp, Fadin:1998py, Meyer:2004jc, Morningstar:1999rf, Chen:2005mg, Lucini:2001ej, 
 Gregory:2012hu, Sun:2017ipk, Athenodorou:2020ani}. A similar analysis within the HW was done for  odd spin glueballs comparing their Regge trajectories with the odderon \cite{FolcoCapossoli:2013eao}, and also for other hadrons with different spins \cite{deTeramond:2005su, Erlich:2005qh}.   

A well known drawback of the HW  model is that it leads to non-linear Regge trajectories. This problem is overcome by the soft wall model which has exact linear Regge trajectories \cite{Karch:2006pv}. This works very well for scalar and vector mesons \cite{Karch:2006pv, Colangelo:2008us} also reproducing masses of light states, but for glueballs, despite the linear trajectory, the mass spectra \cite{Colangelo:2007pt} is not in agreement with lattice data or other approaches. See Tables \ref{Tab1} and \ref{Tab2}, for comparisons. The SW model can be improved in many different ways and in particular if one considers dynamical corrections and anomalous dimensions the glueball spectra becomes good (see Table \ref{Tab2}), but it no longer has linear Regge trajectories \cite{FolcoCapossoli:2015jnm, FolcoCapossoli:2016uns}.

In this work, we consider the inclusion of anomalous dimensions in the conformal dimension of boundary operators in the holographic HW  model. As is well known \cite{Gross:2022hyw}, anomalous dimensions appear in QCD loop corrections, as in the BFKL pomeron \cite{Fadin:1998py}, as well as in a semi-classical limit of gauge/string dualities \cite{Gubser:2002tv}. As we show here, the introduction of anomalous dimensions in the HW model lead to improvements of the model allowing to match the Regge trajectory of the pomeron, $\alpha(t=M^2)=1.08 + 0.25 M^2$, also obtain good glueball masses when compared with  lattice data, and better than the usual HW model. In particular, we show that considering the dimension $\Delta$ of the spin $S$ operators as $\Delta=a\sqrt{S}+b$, where $a$ and $b$
are constants, implies  asymptotic linear Regge trajectories associated with even glueballs. 

This work is organized as follows. 
In section \ref{review}, we briefly review the AdS/CFT correspondence 
and the description of scalar fields in AdS space. In section \ref{OHWModel} we review the main properties of the original HW model which are relevant to the discussion of the anomalous HW model introduced in section \ref{secanomdim}, with
 specific expressions for the anomalous dimensions starting with the logarithm case, truncated series, as a square root and a linear anomalous HW model.  
In section \ref{PomeronAHW}, we give the basic facts of the pomeron Regge trajectory and present detailed discussion of the models introduced in section  \ref{secanomdim}. In section \ref{subseclog}, we present three different fits for the logarithm anomalous dimensions adjusting the pomeron trajectory and predicting glueball masses compared with lattice data, and a fourth case fitting glueball masses and giving the Regge trajectory as an output. In section \ref{subsecapprox}, we present a series expansion of the logarithm anomalous dimensions an truncated approximations for it which we fit with the pomeron trajectory reproducing glueball masses. In section \ref{subsecsqrt}, we present an even simpler expression for the anomalous dimension as a linear plus a square root of the spin of the glueball operator targeting the pomeron trajectory giving very good masses for glueball when compared with lattice data. In section \ref{secLHW}, inspired by the previous cases, we introduce an asymptotic linear anomalous HW model fitting very well the pomeron trajectory and present good glueball masses compared with lattice results. Finally, in section \ref{conclusions}, we present a summary of our results and our conclusions.

\begin{table*}[h]
\caption{\label{Tab1}Masses in GeV of the $J^{++}$ glueball operators with even spins from  lattice data 
 \cite{Meyer:2004jc, Morningstar:1999rf, Chen:2005mg, Lucini:2001ej, 
 Gregory:2012hu, 
 Sun:2017ipk, Athenodorou:2020ani}. In the last column we present average values. The numbers in  parenthesis represent the uncertainties.} 
 \bigskip 
\begin{ruledtabular}
\tiny 
\begin{tabular}
{||c||c|c|c|c|c|c|c|c|c|c|c|c||}
$J^{PC}$ & \cite{Meyer:2004jc}  & \cite{Morningstar:1999rf} & \cite{Chen:2005mg} &  \cite{Lucini:2001ej} &   \cite{Lucini:2001ej} & 
 \cite{Gregory:2012hu} & 
 \cite{Sun:2017ipk} &
\cite{Sun:2017ipk} 
& \cite{Athenodorou:2020ani}
& Average 
\\ \hline
 $0^{++}$ &  1.475(30)(65) & 1.73(05)(08) & 1.71(05)(08) & 1.58(04) & 1.48(07) & 1.795(60) & 
 1.417(30) & 1.498(58) 
 &
 1.653(26)  & 1.59(07) 
 \\
 $2^{++}$ & 2.15(03)(10) & 2.400(25)(120) & 2.39(03)(12) &   &   &  2.620(50) & 2.363(39) & 
 2.384(67) & 
 2.376(32) 
 & 2.38(09)
 \\
 $4^{++}$ & 3.64(09)(16) &  &  &  & & &  & &
 3.69(08)  &3.67(17)
 \\
 $6^{++}$ &  4.36(26)(20) &  &  &  & & &  & & & 
 4.36(46) 
\end{tabular}
\end{ruledtabular}
\end{table*}

\section{Brief review of AdS/CFT and Scalar fields in AdS Space}
\label{review}

Essentially, the AdS/CFT correspondence \cite{Aharony:1999ti} is an equivalence between a superstring theory, formulated in a 10-dimensional spacetime, the $AdS_5\times S^5$, and a super Yang-Mills (SYM) field theory with conformal symmetry, which lives on the boundary of $AdS_5$, the 4-dimensional Minkowski space. Here, $AdS_5$ is the 5-dimensional Anti-de Sitter spacetime, and $S^5$ the sphere in 5 dimensions. The space generated by this product of two manifolds is understood as follows: each point of $AdS_5$ is tangential to the sphere $S^5$.

    A conformal theory in $d$ spacetime dimensions is invariant by a set of transformations characterized by the group $SO(2,d)$, including scale transformations as a particular case.  Essentially, such transformations do not preserve lengths, but preserves angles.

    Anti-de Sitter space is a space of constant negative curvature, while the sphere is a space of constant positive curvature.  When inserted into a higher dimensional (pseudo)-Euclidean space, we can easily describe how its coordinates behave. Let $X^\mu$ be the coordinates of $AdS_5$, with $\mu\in(0,1,...,5)$, and $Y^a$ be the coordinates of $S^5$, with $ a\in(1,2,...,6)$. Such coordinates satisfy the following relations:
    \begin{eqnarray}
        -(X^0)^2+(X^1)^2+(X^2)^2+(X^3)^2+(X^4)^2-(X^5)^2=- R^2,
\cr 
        (Y^1)^2+(Y^2)^2+(Y^3)^2+(Y^4)^2+(Y^5)^2+(Y^6)^2=R^2, 
    \end{eqnarray}
    where $R$ is a constant which plays the role of the radius of both the AdS$_5$ and the $S^5$ spaces. 
    
  Our aim here is to motivate the correspondence based on symmetry arguments. From the point of view of string theory, we have a 10-dimensional spacetime, formed by the direct product of two manifolds. The 5-dimensional Anti-de Sitter space has isometry described by the group of conformal transformations $SO(4,2)$. The 5-dimensional sphere has $SO(6)$ as its isometry group.
    
    The boundary theory living in Minkowski space has exactly the same isometries as the bulk theory with symmetry group $SO(4,2)\otimes SO(6)$. Then, the Minkowski field theory has conformal symmetry  $SO(4,2)$ and it is supersymmetric with 
     $\mathcal{N}=4$ supercharges. As a consequence, it presents $R$ symmetry associated with the group $SU(4)$ which is isomorphic to $SO(6)$. 
    
    Based on these arguments, we see that the symmetries that arise in the two theories are the same. This is not enough to prove correspondence, but it is a strong indication that there is, in fact, a relationship between the two theories.

    In fact, in 1973, Gerard 't Hooft presented  \cite{'tHooft:1973jz} an approximate method to treat $SU(N)$ gauge theories with large $N$, with $N$ being the number of colors in the theory. In this regime, the topological structures of the Feynman diagrams of field theory are identical to the topological structures of the diagrams in string theory. The 't Hooft parameter is given by
    \begin{equation}\label{stringcoupling}
        \lambda=g_{\rm YM}^2 N,
    \end{equation}
    where $g_{\rm YM}$ is the Yang-Mills field theory coupling constant. Therefore, this parameter is directly related to the magnitude of the interaction. Later, the AdS/CFT correspondence would state that
    \begin{equation}\label{relation}
        \lambda=\frac{R^4}{\alpha'^2},
    \end{equation}
    where $R$ is the radius of curvature of $S^5$ and $\alpha'$ is the slope parameter of string theory. Thus, we notice that for a fixed radius $R$, when we have small $\alpha'$, we have large $g_{\rm YM}$, and vice-versa. In this way, we have already noticed the power of correspondence, because, in non-perturbative regimes of field theory, where $g_{\rm YM}$ is large, string theory has small $\alpha'$, being weakly coupled and, therefore, easily treatable. Therefore, we can obtain information from a field theory in its non-perturbative regime through this duality.  
    
The claim that a 10-dimensional string theory is dual to a 4-dimensional field theory may initially seem strange. The idea is similar to holography, where a three-dimensional image is encoded entirely into a two-dimensional object.  The Minkowski spacetime, being the boundary of the Anti-de Sitter space, contains information from theories that propagate in $AdS_5$.
    
    From the initial topics of modern physics, we know wave-particle duality. We can ask ourselves what the fundamental nature is, wave or particle. Today, we know that the answer is none of them, and what is understood is that both wave behavior and corpuscular behavior are classical limits of the theory, the Hilbert space of either description is the same. In the case of the AdS/CFT correspondence, the same occurs: the Hilbert space of string theory in $AdS_5\times S^5$ is the same as that of conformal field theory in 4-dimensional Minkowski.
    
    Now, let's analyze a result that will be useful to us soon, the behavior of a scalar field in $AdS_{d+1}$ \cite{Ramallo:2013bua}. Let's start by writing the metric of this space,
    \begin{equation}\label{metric}
        ds^2=\frac{R^2}{z^2}(dz^2+\eta_{\mu\nu}dx^\mu dx^\nu).
    \end{equation}
    where $R$ is the AdS radius and $\eta_{\mu\nu}$ is the metric of the  Minkowski space with coordinates $x^\mu=(x^0, x^1, \cdots, x^{d-1})$ defined on the $d$ dimensional boundary of the AdS space. So, the conformal border is obtained when we have $z=0$, plus an additional point at $z\to \infty$. 
    
    A scalar field $\phi$ with mass $m$ in the space defined by the metric (\ref{metric}) is governed by the Klein-Gordon equation:
    \begin{equation}\label{kgequation1}
        z^{d+1}\partial_z(z^{1-d}\partial_z\phi)+z^2\eta^{\mu\nu}\partial_\mu\partial_\nu\phi-(mR)^ 2\phi=0.
    \end{equation}
    
    Let us then perform a Fourier transform of the field $\phi$ in the coordinates $x^\mu$, that is,
    \begin{equation}
        \phi(z,x^\mu)=\int\frac{d^dk}{(2\pi)^d}e^{ik\cdot x}f_k(z).
    \end{equation}

In this way, the equation of motion becomes
    \begin{equation}
        z^{d+1}\partial_z(z^{1-d}\partial_zf_k)-k^2z^2f_k-(mR)^2f_k=0.
    \end{equation}
    
    We are interested in its behavior near the frontier $z=0$. If we impose such a condition on the equation above, we note that a natural solution is $f_k(z)\sim z^\beta$, with $\beta$ satisfying 
    \begin{equation}
        \beta(\beta-d)-(mR)^2=0,
    \end{equation}
    so that 
    \begin{equation}
        \beta=\frac{d}{2}\pm\sqrt{\frac{d^2}{4}+(mR)^2}.
    \end{equation}
    
    Defining
    \begin{equation}\label{massdim}
        \Delta\equiv\frac{d}{2}+\sqrt{\frac{d^2}{4}+(mR)^2},
    \end{equation}
    we have that, close to the border $z\sim0$, the function $f_k(z)$ behaves as
    \begin{equation}
        f_k(z)\approx A(k)\, z^{d-\Delta}+B(k)\, z^{\Delta}.
    \end{equation}
    
    We can then perform an inverse Fourier transform, in order to obtain the field close to the boundary ($z\rightarrow0$) in the configuration space:
    \begin{equation}
        \phi(z,x)\approx A(x)\, z^{d-\Delta}+B(x)\, z^\Delta.
    \end{equation}
    
    It is important to note that $\Delta$ is real if the term inside the root is greater than or equal to zero, that is,
    \begin{equation}
        m^2\geq-\bigg(\frac{d}{2R}\bigg)^2,
    \end{equation}
which is known as the Breitenlohner-Freedman bound \cite{Breitenlohner:1982jf}.

    If the above condition is valid, we see that $d-\Delta\leq\Delta$. Thus, the term $z^{d-\Delta}$ is dominant as we approach $z=0$. In order to have a finite field operator $\varphi(x)$ on the border we write
    \begin{equation}
    \varphi(x)=\lim_{z\rightarrow0}z^{\Delta-d}\phi(z,x). 
    \end{equation}
The corresponding action coupling $\phi(z,x)$ to a boundary operator ${\cal O}$ evaluated at $z\to \epsilon$ is the following
    \begin{equation}
    S_{\rm boundary} \sim \int d^dx \, \sqrt{\gamma_\epsilon} \, \phi(\epsilon,x)\,{\cal O}(\epsilon,x)\,,
    \end{equation}
where $\gamma_\epsilon=(R/\epsilon)^{2d}$ is the determinant of the induced metric at $z=\epsilon$. 
Then, 
    \begin{equation}
    S_{\rm boundary}  \sim R^d \, \int d^dx \,  \varphi(x) \, \epsilon^{-\Delta}  
    \,{\cal O}(\epsilon,x)\,.
    \end{equation}
To have a finite and $\epsilon$-independent boundary action we define
    \begin{equation}
    {\cal O}(\epsilon, x) \equiv \epsilon^{\Delta}  
    \,{\cal O}(x)\,. 
    \end{equation}
 The above relationship shows us, therefore, that scalar excitations of the string, of mass $m$, couple to field operators on the boundary that have dimensions $\Delta$. This result is extremely important in the analysis we will carry out to calculate hadron masses in the following.

\section{The original HW  model}
\label{OHWModel}

   The AdS/CFT correspondence alone will not allow us to calculate the masses of glueballs, once in a CFT physical quantities are massless. In the HW  model \cite{Boschi-Filho:2002xih, Boschi-Filho:2002wdj} one considers the AdS metric given by Eq. 
\eqref{metric} and introduces a cut in the  holographic $z$ coordinate imposing that $z\in[0,z_{\rm max}]$, such that 
    \begin{equation}\label{zmax}
        z_{\rm max}=\frac{1}{\Lambda_{\rm QCD}},
    \end{equation}
where $\Lambda_{\rm QCD}$ is a typical QCD mass scale with a value around 150-300 MeV.  

As discussed in the previous section, from the AdS/CFT correspondence one is able to show that scalar excitations with mass $m$ in AdS couple to scalar field operators at the boundary whose dimensionality is $\Delta$, given by Eq. (\ref{massdim}). Still, according to the correspondence a spin $J>0$ field in AdS is  equivalent to a massless spin $J$ field operator on its border \cite{Aharony:1999ti}.

Phenomenologically, it was proposed that in the HW  model the glueball operator with dimension $\Delta$ and nonzero spin couples to a massive scalar excitation in AdS according to \cite{deTeramond:2005su} 
    \begin{equation}\label{delta}
        \Delta=2+\sqrt{4+(mR)^2}, 
    \end{equation}
    or
    \begin{equation}\label{mR2}
        (mR)^2=\Delta (\Delta + 4).  
    \end{equation}
    The scalar glueball operator on the boundary is given by ${\cal O}_4={\rm tr} {F_{\mu\nu}}^a {F^{\mu\nu}}^a\equiv {\rm tr} F^2$ which has 
    mass dimension $\Delta=4$ and $F_{\mu\nu}^a$ is the usual Yang-Mills tensor with non-Abelian index $a\in(1, \dots, N)$ of the $SU(N)$ gauge group. Following the idea \cite{Gubser:2002tv} that high spin  operators can be constructed by inserting covariant derivatives $D_\mu$ into lower spin operators one can write high spin $S$ glueball operators as 
    \begin{equation}\label{highS}
    {\cal O}_{4 + S}\,=\, {\rm tr} F D_{\{\mu_1}...D_{\mu_S\,\}} F,
    \end{equation}
    such that in four dimensions they have conformal dimension 
     $\Delta=4+S$ \cite{deTeramond:2005su}. Then, one has
    \begin{equation}\label{spinmass}
    \begin{split}
        &4+S=2+\sqrt{4+(mR)^2}\\
        &(S+2)^2=4+(mR)^2\\
        &(mR)^2=S(S+4).
    \end{split}
    \end{equation} 
This is a twist $\tau=4$ tower of operators, since in general the twist is defined as $\tau=\Delta-S$. 
It is important to mention that these mass dimensions are all canonical: 
\begin{equation}\label{deltacan}
    \Delta_{\rm can.}=4+S. 
\end{equation}
 Therefore,  the equation of motion associated with glueballs in AdS$_5$ space, Eq. (\ref{kgequation1}),  becomes
    \begin{equation}\label{eomAdS5}
        \bigg[z^3\partial_z\frac{1}{z^3}\partial_z+\eta^{\mu\nu}\partial_\mu\partial_\nu-\frac{(mR)^2}{z^ 2}\bigg]\phi=0.
    \end{equation}
    Substituting the relation (\ref{spinmass}) in the above equation, one has 
    \begin{equation}
        \bigg[z^3\partial_z\frac{1}{z^3}\partial_z+\eta^{\mu\nu}\partial_\mu\partial_\nu-\frac{S(S+4)}{z ^2}\bigg]\phi=0.
    \end{equation}
    Taking a plane wave ansatz in $\vec{x}$ and in time $t$, that is,
    \begin{equation}
        \phi(x,z)=Ae^{-iP\cdot x}f(z),
    \end{equation}
    and, substituting it into the equation of motion, one finds 
    \begin{equation}\label{solutionphi}
        \phi(x,z)=C_{\nu,k}\, e^{-iP\cdot x}z^2J_\nu(u_{\nu,k}z),
    \end{equation}
    where $C_{\nu,k}$ are normalization constants that will not be important in our analysis, $J_\nu(u_{\nu,k}z)$ is the Bessel function of order  $\nu=2+S$, $u_{\nu,k}$ $(k=1,2,...)$ are discrete modes determined by boundary conditions. In this work we will impose Dirichlet boundary conditions, that is,
    \begin{equation}
        \phi (x, z)|_{\,z =z_{\rm max}}= 0, 
    \end{equation}
    which implies that 
    \begin{equation}
\label{Dmasses}
u_{\nu , k} \,=\, \frac{\chi_{_{\nu , k}}} {z_{\rm max}}\,=\,\chi_{_{\nu , k}}\,\Lambda_{\rm QCD}
\qquad ;\qquad J_\nu (\,\chi_{_{\nu , k}}\,) \,=\,0\,. 
\end{equation}
  The scale $\Lambda_{\rm QCD}$ is usually fixed using some experimental or lattice data.  The $k$ indices label the radial excitations of the particle states with masses proportional to the zeros of the Bessel function $J_\nu (w)$, {\sl i.e.}, proportional to the value of the argument $w$ where the corresponding Bessel function is zero.

So, using the above discussion of the HW model, one can calculate higher spin glueball masses from relations \eqref{Dmasses} and \eqref{deltacan} such that 
with $\nu=\Delta-2=S+2$
    \begin{equation}\label{relation1}
        u_{S+2,k}=\chi_{S+2,k}\, \Lambda_{\rm QCD}. 
    \end{equation}
 Using this relation, one finds the glueball masses $M_i$ from the original HW \cite{Boschi-Filho:2005xct}, 
    \begin{eqnarray}\label{massratioHW}
        \frac{M_0}{\chi_{2,1}}=\frac{M_2}{\chi_{4,1}}
        =\frac{M_4}{\chi_{6,1}}=\cdots
        =\frac{M_i}{\chi_{i+2,1}}=\cdots , 
    \end{eqnarray}
 which are presented here for latter convenience in Table \ref{Tab2}, and the corresponding relative deviations  $\delta_{i}=|M_{i}-M_{\rm latt}|/M_{\rm latt}$ with respect to the average of lattice data presented in Table \ref{Tab1}.
 We use the mass of the $0^{++}$ glueball state from lattice as an input. For comparison, we also include the corresponding masses calculated from the SW model given by $m_J^2=k\left[4+ 2\sqrt{4+J(J+4)} \right]=2k(J+4)$ with $k_1=0.316$ GeV$^2$ (SW1) and $k_2=1.00$ GeV$^2$ (SW2) \cite{FolcoCapossoli:2015jnm}, and the dynamical anomalous SW (DASW) model \cite{FolcoCapossoli:2016uns}. Despite the SW model provides a linear trajectory, as we can see, the masses obtained are not in good agreement with the masses obtained by lattice QCD. In the case where the masses presents small deviations (DASW), the Regge trajectory becomes non-linear. This is a good reason to search for a possibility of linearity in HW model, since it could provide good masses combined with a linear Regge trajectory.

\begin{table*}
\caption{\label{Tab2}Masses in GeV of the $J^{++}$ glueball operators with even spins from the original  HW, the original SW  (SW1 and SW2), and the dynamical anomalous SW   (DASW) models, with the 
corresponding deviations from the average lattice data presented in Table I. See the text for details.  
The masses of the $0^{++}$ state are inserted as inputs for the different models.} 
\begin{ruledtabular}
\begin{tabular}{||c||cc|cc|cc|cc||}
$J^{PC}$  & HW & $\delta_{\rm HW}\;\;\quad$ &  SW$_1$ & $\delta_{\rm SW1}\;\;\quad $ &  SW$_2$ & $\delta_{\rm SW2}\;\;\quad $ &  DASW & $\delta_{\rm DASW}\;\;\quad $
\\ \hline
 $0^{++}$ & $1.59$ & $0.0\%\;\;\quad$ & $1.59$ & $0.0\%\;\;\quad$ & $2.83$ & $78\%\;\;\quad$ & $1.56$ & $1.9\%\;\;\quad$ 
 \\
 $2^{++}$ &$2.35$&  $1.3\%\;\;\quad$  &$1.95$&  $18\%\;\;\quad$ &$3.46$&  $45\%\;\;\quad$ &$2.52$&  $5.9\%\;\;\quad$
 \\
 $4^{++}$ &$3.08$&  $16\%\;\;\quad$  &$2.24$&  $39\%\;\;\quad$ &$4.00$&  $9.0\%\;\;\quad$ &$3.43$&  $6.5\%\;\;\quad$ 
 \\
 $6^{++}$ &$3.78$&  $13\%\;\;\quad$   &$2.51$&  $42\%\;\;\quad$ &$4.47$&  $2.5\%\;\;\quad$ &$4.32$&  $0.9\%\;\;\quad$ 
 \\
 $8^{++}$& $4.48$&   - \;\;\quad & $2.75$&  - \;\;\quad& $4.90$&   - \;\;\quad & $5.19$&   -  \;\;\quad 
 \\
 $10^{++}$&  $5.17$&   -  \;\;\quad&$2.97$&   -  \;\;\quad&$5.29$&   -   \;\;\quad&$6.05$&   -    \;\;\quad
\end{tabular}
\end{ruledtabular}
\end{table*}


 \section{Anomalous HW  Models}
\label{secanomdim}

Before we introduce the anomalous HW  model, let us briefly discuss the role of anomalous dimensions in quantum field theory. In a scale invariant field theory, such as a CFT, under a dilation $x\rightarrow\lambda x$ the operators $\mathcal{O}$ by definition behave as
\begin{equation}
    \mathcal{O}\rightarrow\lambda^{-\Delta}\mathcal{O}, 
\end{equation}
where $\Delta$ is the conformal dimension of $\mathcal{O}$. In free field theories $\Delta$ is simply the mass dimension of the operator that can be read off directly from the lagrangian by dimensional analysis. Nonetheless, for interacting field theories the renormalization process modifies this dimension according to 
\begin{equation}
    \Delta=\Delta_{\rm can.} + \gamma(g),
\end{equation}
where $\Delta_{\rm can.}$ is the canonical conformal dimension of the operator in a free field theory, $\gamma(g)$ is the so-called anomalous dimension, and $g$ is the running coupling of the theory. So, the anomalous dimension gives us a measure of the deviation of the conformal dimension from the value that it would assume in a free field theory.

It is clear that anomalous dimension becomes relevant when there is interaction in the theory. Therefore, it may not be obvious how it can emerge in the HW model, since for $z<z_{\rm max}$ the coupling constant seems to be null. Despite the fact that we are working with free theories in the AdS bulk, it is an approximation. As we can see in equations (\ref{stringcoupling}) and (\ref{relation}), the string theory coupling constant $\alpha'$ and the coupling constant of the field theory in the boundary, $g_{\rm YM}$, have a relation: when the first one becomes small, the second becomes big and vice versa. We are interested of treating the non-perturbative regime of QCD at the boundary (large $g_{\rm YM}$), that implies a small (that we treat as zero in first approximation) $\alpha'$ at the bulk.

Even if the constant $\alpha'$ is treated as null, the operators that we are interested in have anomalous dimensions. The glueball operators, as evidenced by Eq. (\ref{highS}), are formed by $F_{\mu\nu}^a$ and these last by elementary operators $A_{\mu}^a$, that describes the $SU(N)$ gauge field. These are the operators that have their conformal dimension changed by interaction. At the boundary, the coupling constant of the theory $g_{\rm YM}$ assumes a finite value that implies  an  anomalous dimension for $F_{\mu\nu}^a$ and consequently, for the glueball operators.


    Our proposal here, is to include anomalous dimensions in the conformal dimension of the boundary glueball operators, modifying the relation between $\Delta$ and $S$. 
     It is well known that anomalous dimensions play a important role in the renormalization of QCD (see {\sl e.g.} \cite{Gross:2022hyw}) and the BFKL pomeron \cite{Fadin:1998py}, however it is difficult to relate it to high spin states in field theory.

     On the other hand, from a semi-classical limit of gauge/string dualities \cite{Gubser:2002tv}, one finds that the conformal dimension of  dual operators with spins $S$, as the ones given by Eq. \eqref{highS}, behaves  differently as one increases the spin $S$ in comparison with the square root of the 't Hooft coupling $\lambda$. Basically, one finds three regimes:
     \begin{itemize}
     
     \item If the spins are small which means $S \ll \sqrt{\lambda}$, then the operators $\mathcal{O}$ have canonical conformal dimension 
     \begin{equation}\label{Delta4+S}
     \Delta = 4+S, 
     \end{equation}
     as in Eq. \eqref{deltacan}; 
     
     \item If the spins are large as $S \gg \sqrt{\lambda}$, then 
     \begin{equation}\label{DeltaS}
        \Delta=S + \frac{\sqrt{\lambda}}{\pi}\ln \left(\frac{S}{\sqrt{\lambda}} \right)
        +{\cal{O}}(S^0),  
\end{equation}
        so that the anomalous dimensions are given by
     \begin{equation}\label{Delta_anom}
        \Delta_{\rm anom.}= \frac{\sqrt{\lambda}}{\pi}\ln \left(\frac{S}{\sqrt{\lambda}} \right)
        +{\cal{O}}(S^0); 
\end{equation}
     \item  If the spins are of the same order of the square root of the 't Hooft coupling $ S \sim  \sqrt{\lambda}$, then some other unkown complicated non-perturbative relation between $\Delta$ and $S$ should hold. 
     \end{itemize}

The phenomenological anomalous holographic HW  model that we are proposing starts with the hard cutoff, Eq. \eqref{zmax}, the equation of motion, Eq. \eqref{eomAdS5}, and the higher spin glueball masses are 
given by 
    \begin{eqnarray}\label{massratioAHW}
        \frac{M_0}{\chi_{2,1}}=\frac{M_i}{\chi_{\nu_i,1}},
    \end{eqnarray}
with the index of the Bessel function is given by $\nu_i=\Delta_{\rm AHW_i}-2$, where $\Delta_{\rm AHW_i}$ takes into account the anomalous dimensions discussed above. The uncertainties in the masses are then:
    \begin{eqnarray}\label{uncert}
        \delta{M_i}=\sqrt{\left(\frac{\chi_{\nu_i,1}}{\chi_{2,1}}\right)^2 \left(\delta M_0\right)^2 + \left(\frac{M_0}{\chi_{2,1}}\right)^2 
        \left(\delta \chi_{\nu_i,1}\right)^2}, 
    \end{eqnarray}
where the uncertainties $\delta M_0$ come from lattice data and $\delta\chi_{\nu_i,1}$ from the determination of the zero of the Bessel functions for high spins due to uncertainties in the anomalous dimensions to be discussed below.  
For low spins, we suppose the canonical dimension, Eq. \eqref{Delta4+S}, holds, in which case $\delta \chi_{\nu_i,1}=0$, while for high spins the anomalous dimensions, Eq. \eqref{Delta_anom}, are assumed without an intermediate regime as follows: 
\begin{eqnarray}
\Delta_{\rm AHWlog}= \left\{
\begin{array}{lccccc}
4+S; &&&&& 0 \le S \le S_0; \\ 
S + a \ln \left({S} \right) + b; &&&&&  S > S_0,
\end{array}
\right. 
\end{eqnarray}
where the value of the spin $S_0$ of the transition between low ($0 \le S \le S_0$) and high ($S > S_0$) spins and the constants $a$ and $b$ will be determined by the best fit to experimental and lattice data. As we are going to see below in section \ref{subseclog}, 
this model produces interesting results, improving the ones from the usual HW and soft wall models regarding the predicted masses and from the HW  with respect to the corresponding Regge trajectories. 

Inspired by these results, we also consider other phenomenological anomalous HW models where we introduce some approximations for the anomalous dimensions for high spins. The first approximation we consider is motivated by truncated series expansions of the relation  $\ln x = 2 \arcsinh \left[(1/2)\left( \sqrt{x} - {1}/{\sqrt{x}}\right)\right] $ leading to the models: 
\begin{eqnarray}
\Delta_{\rm ATSHW}= \left\{
\begin{array}{lccccc}
4+S; &&&&& 0 \le S \le S_0; \\ 
 S + a \sum_{k=0}^N 
        \frac{(-1)^k}{2^{2k}(2k+1)k!} 
        \left(\frac 12 \right)_k 
    \left(  \sqrt{S} - \frac{1}{\sqrt{S}}
    \right)^{2k+1}  
        +b; &&&&&  S > S_0,
\end{array}
\right. 
\end{eqnarray}
with $N=0, 1, 2, 3$, discussed in Section \ref{subsecapprox} from which we obtain good glueball masses and Regge trajectories. 

Further, we consider another approximation leading to an AHW model with a square root anomalous dimension for high spins:
\begin{eqnarray}
\Delta_{\rm AHWSQRT}= \left\{
\begin{array}{lccccc}
4+S; &&&&& 0 \le S \le S_0; \\ 
 S+ a\sqrt{S} +b\,; &&&&&  S > S_0,
\end{array}
\right. 
\end{eqnarray}
presented in Section \ref{subsecsqrt}, which give very good results for glueball masses although not so good Regge trajectories. 

Finally, we consider an even simpler expression for the anomalous dimension for high spins without the linear term $S$: 
\begin{eqnarray}
\Delta_{\rm ALHW}= \left\{
\begin{array}{lccccc}
4+S; &&&&& 0 \le S \le S_0; \\ 
 a \sqrt{S}+b\,.; &&&&&  S > S_0,
\end{array}
\right. 
\end{eqnarray}
discussed in Section \ref{secLHW} which leads to good glueball masses and {\sl linear} Regge trajectories in this AHW model.  


\section{Glueballs and the Soft Pomeron in The AHW models}\label{Glueball+pomeron}
\label{PomeronAHW}

The rise of the proton-proton cross section with energy is related to the soft pomeron, a particle with no charges and quantum numbers of the vacuum, whose experimental Regge trajectory is the following
\cite{Landshoff:2001pp, Meyer:2004jc}
\begin{equation}\label{pomeron}
  \alpha(t=M^2) =  1.08 + 0.25 M^2  ,
\end{equation}
with masses $M$ expressed in GeV. The BFKL pomeron \cite{Fadin:1998py} is a perturbative treatment of the problem, although non-perturbative aspects are also present (see {\sl e.g.} \cite{Lebiedowicz:2023mhe}).

In this work we analyse many possibilities of anomalous conformal dimensions for glueball operators and compare their masses with the masses of average lattice, presented in Table \ref{Tab1}. Despite we could use the best match with such masses of lattice as the criterion to determine the best model we are interested in a connection with the pomeron. 

For that reason, our criterion of best fit is defined as follows: in a model, a given anomalous dimension produces some values for glueball masses. With these masses we use linear regression to obtain the corresponding Regge trajectory. The best fit is obtained when such Regge trajectory is as close as possible to the soft pomeron, given by Eq. \ref{pomeron}, with smaller $\chi^2$/ndf.  

In all models considered in this work, we assume that the state $0^{++}$ has canonical conformal dimension $\Delta=4$, and use it as an input with mass $M_{0^{++}}=1.59\mbox{ GeV}$ from lattice. Following Landshoff \cite{Landshoff:2001pp}, we take that this state does not belong to the soft pomeron trajectory.


\subsection{Logarithm anomalous dimensions}
\label{subseclog}

\begin{figure}[h]
\vskip0.5cm 
\centering
\includegraphics[width=14cm]{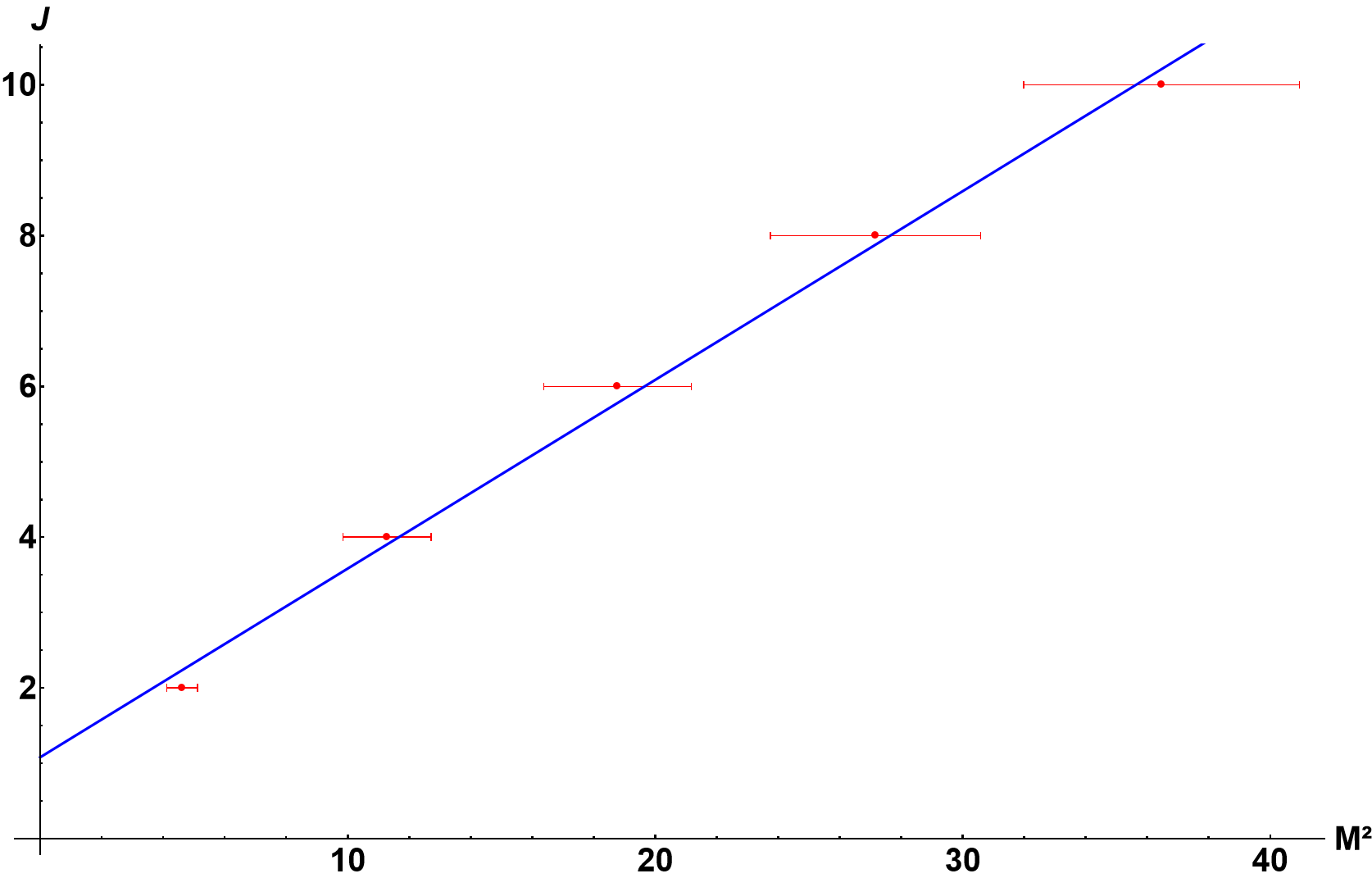}
\hskip0.8cm 
\caption{Plot of $J\times M^2$ with masses expressed in GeV and $J=2$ to $10$ in the AHWlog model with anomalous dimensions given by Eq. \eqref{deltalog1}. The dots represent the glueball masses shown in the 2nd column of Table \ref{tablelog} with the corresponding error bars. The  straight line corresponds to a linear fit of the soft pomeron trajectory  $ J =  1.08 \pm 0.21 + (0.25 \pm 0.01) M^2 $ with $\chi^2$/ndf=3.76/3=1.25.}
\label{image1}
\end{figure}

Here, we start with the AHWlog model 
\begin{eqnarray}
\Delta_{\rm AHWlog}= \left\{
\begin{array}{lccccc}
4+S; &&&&& 0 \le S \le S_0; \\ 
\Delta_{\rm log}; &&&&&  S > S_0,
\end{array}
\right. 
\end{eqnarray}
where 
\begin{eqnarray}\label{deltalogab}
\Delta_{\rm log}= 
S + a \ln \left({S} \right) + b, 
\end{eqnarray}
and the constants $S_0$, $a$, and $b$  will determined fitting experimental and lattice data below. First, we consider that $0\le S \le 10$ and $S_0=0$. The best fit obtained in this way implies the coefficients $a=1.92\pm0.36$, and $b=2.13\pm0.07$, such that the effective dimension in this anomalous HW model is given by
\begin{eqnarray}\label{deltalog1}
\Delta_{\rm log1}= 
S +(1.92\pm0.36)\ln{(S)}+2.13\pm0.07;  \qquad\qquad (2\le S \le 10).
\end{eqnarray}
From this anomalous dimension, we obtain the glueball masses $M_i$ shown in the 2nd column of Table (\ref{tablelog}) with the corresponding errors given by Eq. \eqref{uncert}, together with the deviations $\delta_{i}=|M_{\rm latt.}-M_{i}|/M_{\rm latt.}$ from average lattice data (Table \ref{Tab1}) and the effective anomalous dimensions of the glueball operators $J^{++}$. In Figure~\ref{image1}, we present the Regge trajectory which is build up as a linearization of these glueball masses, reproducing the soft pomeron trajectory, $ J =  1.08 \pm 0.21 + (0.25 \pm 0.01) M^2 $ with $\chi^2$/ndf=3.76/3=1.25. For clarity, we also show in Table (\ref{tablelog}) the order of the Bessel function for each glueball state in each AHWlog model.


\begin{table*}[h]
\caption{\label{tablelog}Masses in GeV of the $J^{++}$ glueball operators from $J=0$ to $J=10$ for the logarithm anomalous  HW  models (HW$_{\rm log1}$) and (HW$_{\rm log2}$) defined by Eq. \eqref{Dmasses} with the logarithm contributions to the anomalous dimensions Eqs. \eqref{deltalog1} and \eqref{deltalog2}, respectively, with errors calculated according to Eq. \eqref{uncert}. 
The orders $\nu$ of the corresponding Bessel functions are shown for each glueball state for the two models. The mass of the $0^{++}$ with uncertainties are inserted as  inputs from lattice.  We also show the relative deviations $\delta_{\rm log1,2}$ compared with lattice data, and the corresponding anomalous dimensions $\Delta_{\rm anom.}^{\rm log1,2}\equiv\Delta_{\rm log1,2} -(4+S)$ of the states $J^{++}$ in this model.}
\vskip0.5cm 
\begin{ruledtabular}
\begin{tabular}{||c||cccc|cccc||}
$J^{PC}$ & $\nu$ & HW$_{\rm log1}$ &   $\delta_{\rm log1}$ & $\Delta_{\rm anom.}^{\rm log1}$  & $\nu$ & HW$_{\rm log2}$ &   $\delta_{\rm log2}$ & $\Delta_{\rm anom.}^{\rm log2}$ 
\\ \hline
 $0^{++}$ & 2 & $1.59\pm0.07$  & $0\%$ & $0.0$ & 2 & $1.59\pm0.07$  & $0\%$ & $0.0$
 \\
 $2^{++}$ & 3.46 &$2.15\pm0.12$ &$9.71\%$ &$-0.54$  & 4 &$2.35\pm0.10$ &$1.26\%$ &$0.0$
 \\
 $4^{++}$ & 6.79 & $3.36\pm0.21$&$8.49\%$ &$0.79$ & 6.50 & $3.25\pm0.24$&$11.35\%$ &$0.50$
 \\
 $6^{++}$  & 9.57 & $4.33\pm0.28$&$0.63\%$ &$1.57$  & 9.44 & $4.29\pm0.29$&$1.69\%$ &$1.44$
 \\
 $8^{++}$ & 12.12 & $5.21\pm0.33$& - &$2.12$ & 12.10 & $5.21\pm0.34$& - &$2.10$
 \\
 $10^{++}$ & 14.55  & $6.04\pm0.37$& - &$2.55$  & 14.62 & $6.06\pm0.37$& - &$2.62$
\end{tabular}
\end{ruledtabular}
\end{table*}

\begin{figure}[h]
\vskip0.5cm 
\centering
\hskip0.8cm 
\includegraphics[width=14cm]{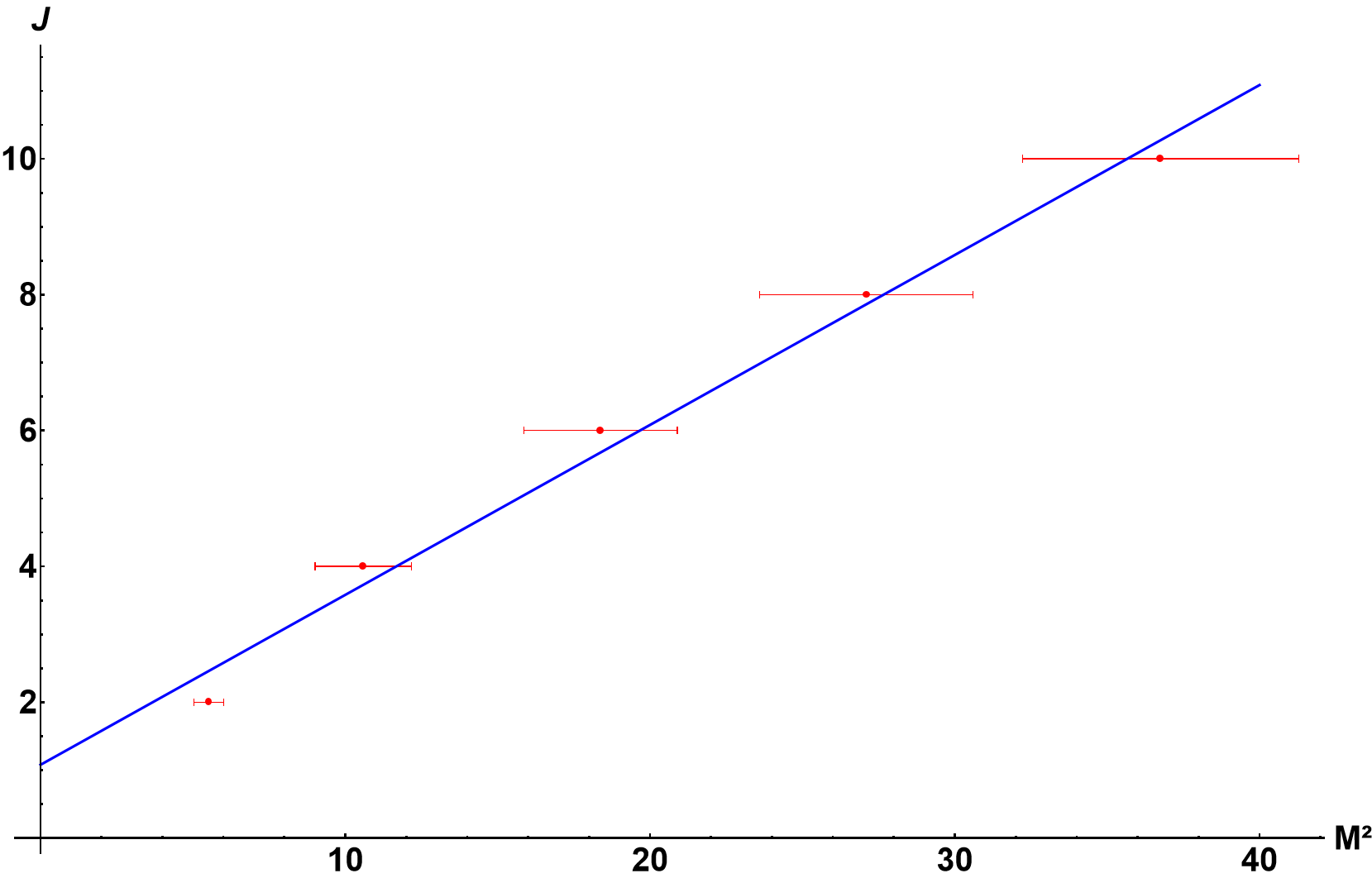}
\caption{Plot of $J\times M^2$ with masses expressed in GeV and $J=2$ to $10$ in the anomalous  model HW$_{\rm log2}$, Eq.  \eqref{deltalog2}. The dots represent the glueball masses shown in the fifth column of Table \ref{tablelog} with the corresponding error bars. The  straight line corresponds to  a linear fit matching the soft pomeron trajectory $ J =  1.08\pm0.36 + (0.25\pm0.02) M^2 $ with $\chi^2/$ndf$=15.2/3=5.1$.}
\label{image2}
\end{figure}

The second case of the AHWlog model that we analyse is the one corresponding to 
$0\le S \le 10$ and $S_0=2$. In this way, the states $0^{++}$ and $2^{++}$ have canonical conformal dimensions $\Delta=4$ and $\Delta=6$, respectively, while higher spin states have anomalous dimensions given by Eq. \eqref{deltalogab}. Applying this model to fit the pomeron trajectory and to obtain glueball masses for the states $J^{++}$ from $J=2$ to $J=10$, we find that the best fit corresponds to
\begin{equation}\label{deltalog2}
\Delta_{\rm log2}= S +(2.32\pm0.25)\ln{(S)}+1.28\pm0.21; \qquad\qquad (4\le S \le 10).
\end{equation}
Within this AHWlog2 model we reobtain the soft pomeron trajectory $ J =  1.08\pm 0.36 + (0.25\pm 0.02) M^2 $, which is shown in Figure \ref{image2}, with $\chi^2/$ndf$=15.2/3=5.1$. The glueball masses are presented in the fifth column of Table \ref{tablelog} with the corresponding errors given by Eq. \eqref{uncert}, together with the deviations with respect to average lattice data and the corresponding anomalous dimensions of the states $J^{++}$ in this model.

Thirdly, we consider the AHWlog model with $0\le S \le 10$ and $S_0=4$, so that the states $0^{++}$, $2^{++}$ and $4^{++}$ have canonical conformal dimensions, whilst higher spins follow Eq. \eqref{deltalogab}. In this case it is not possible the match the soft pomeron trajectory exactly. The closest trajectory found with least $\chi^2$ is $ J =  1.10 \pm 0.47 + (0.27 \pm 0.02) M^2 $, with $\chi^2/$ndf$=24.6/3=8.2$ and effective dimension 
\begin{equation}\label{deltalog3}
\Delta_{\rm log3}=S+(2.34\pm0.04)\ln(S)+0.51\pm0.02; \qquad\qquad (6\le S \le 10), 
\end{equation}
starting with the state $6^{++}$, implying masses $\{2.35\pm0.10, 3.08\pm0.14, 4.06\pm0.18,4.96\pm0.22, 5.82\pm0.26 \}$ GeV for the states from $J=2$ to $J=10$, with relative deviations $\{1.3, 16, 7.5\}\%$ compared with average lattice for the states $2^{++}$, $4^{++}$, and $6^{++}$, respectively.


 A comparison between the above three  fits within the AHWlog model favours the first with respect to second and the third since it gives least $\chi^2$ and $\chi^2$/ndf. In the same token, the second AHWlog is better than the third, thanks to  smaller $\chi^2$ and $\chi^2$/ndf. 


Another possible way to apply the AHWlog model with logarithm anomalous dimensions is to minimize directly the deviations of the glueball masses with respect to average  lattice data and then look up for the resulting Regge trajectory. We consider this case for $0\le S \le 10$ and $S_0=0$, such that the state $0^{++}$ has canonical dimension $\Delta=4$ and the higher spins from $J=S=2$ to 10 have anomalous dimension given by Eq. \eqref{deltalogab}. 
Applying this procedure one finds the coefficients $a=2.3\pm0.8$ and $b=2.5\pm1.1$, which means an effective dimension 
\begin{equation}\label{deltalog4}
\Delta_{\rm log4}= S +(2.3\pm0.8)\ln{(S)}+2.5\pm1.1; \qquad\qquad (2\le S \le 10),
\end{equation}
obtaining exactly the average lattice masses for the $2^{++}$ and $4^{++}$ states besides the input $0^{++}$. For the state $6^{++}$, we find $M_{6^{++}}=4.69$ GeV which is $7.6\%$ higher than the average lattice result. So, this model produces masses with lowest total relative deviation with respect to average lattice outputs when compared with the AHWlog models discussed above. Within this model the predicted soft pomeron trajectory is not so good and is given by $ J =  0.93\pm 0.18 + (0.22 \pm 0.01) M^2 $ with $\chi^2/$ndf$=0.705/3=0.235$. If we want to compare this model with others, we should compute $\chi^2$ based on soft pomeron trajectory instead, by doing this we find $\chi^2/$ndf$=5.62/3=1.87$.

\subsection{Truncated Series for anomalous dimensions} 
\label{subsecapprox}

The conformal dimension for high spins from gauge/string duality discussed in the previous section was developed to describe an ideal situation where the four dimensional field theory is conformal. Since this is not the case of strong interactions that lead to the Regge trajectory of the soft pomeron and to the lattice glueball masses, it is also interesting to investigate some approximate expressions for this quantity since we are studying a phenomenological model, and analyse whether these expressions might give better results compared with experimental and lattice data. 

\begin{figure}[h]
\vskip0.5cm 
\centering
\includegraphics[width=14cm]{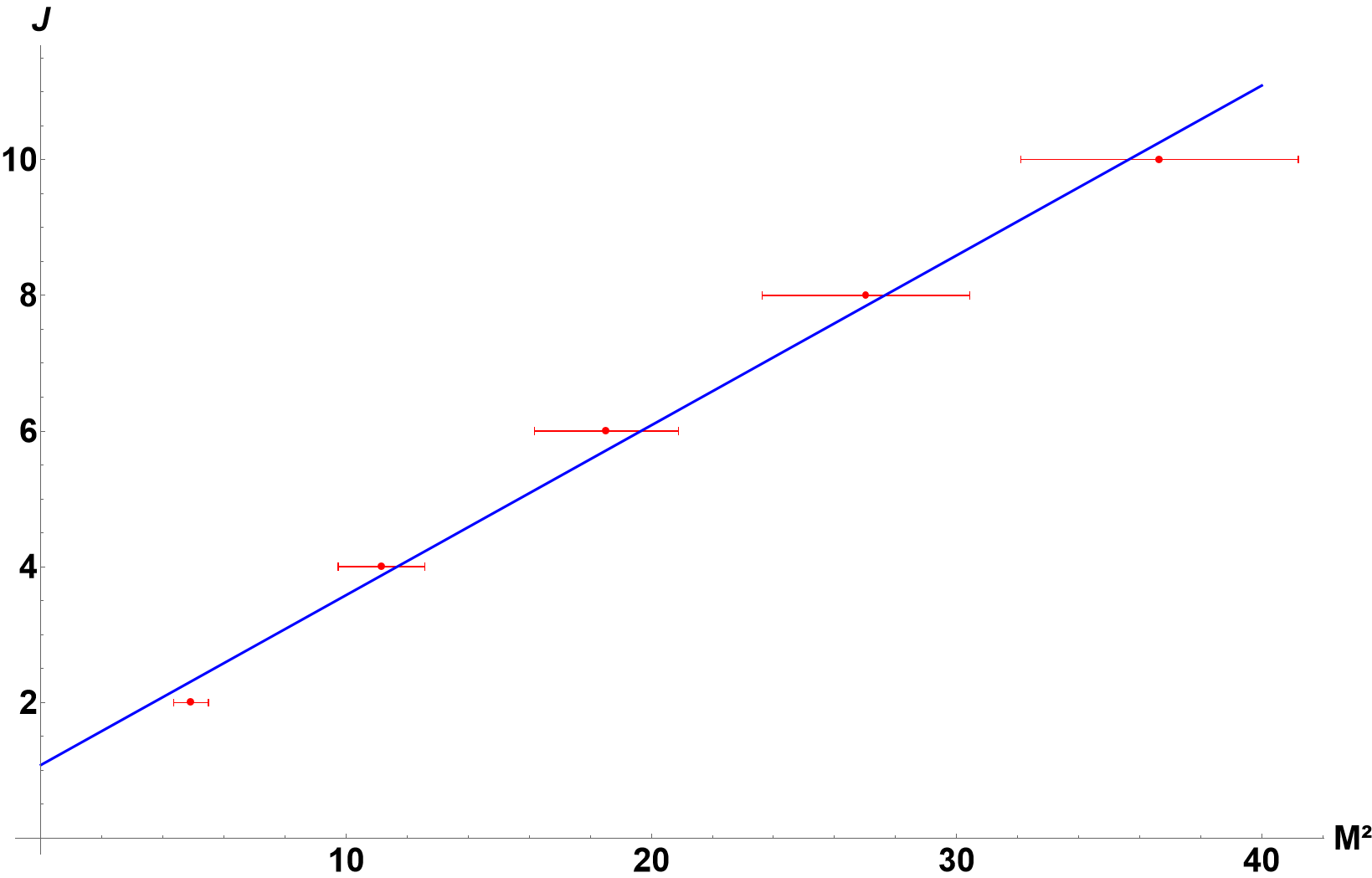}
\hskip0.8cm 
\caption{Plot of $J\times M^2$ with masses $M$ expressed in GeV and $J=2$ to $10$, using the anomalous HW model \eqref{AHWN} with conformal dimension Eq. \eqref{anomalousdim4}, truncated at $N=0$ with coefficients $a=1.38$ and $b=2.67$. The dots with error bars represent the glueball masses shown in Table \ref{tableapprox} for $N=0$, and the straight line corresponds to a linear fit given by $ J =  1.08\pm0.28+ (0.25\pm0.01) M^2 $ with $\chi^2/$ndf$=5.21/3=1.74$.} 
\label{image3}
\end{figure}

\begin{figure}[h]
\vskip0.5cm 
\centering
\hskip0.8cm 
\includegraphics[width=14cm]{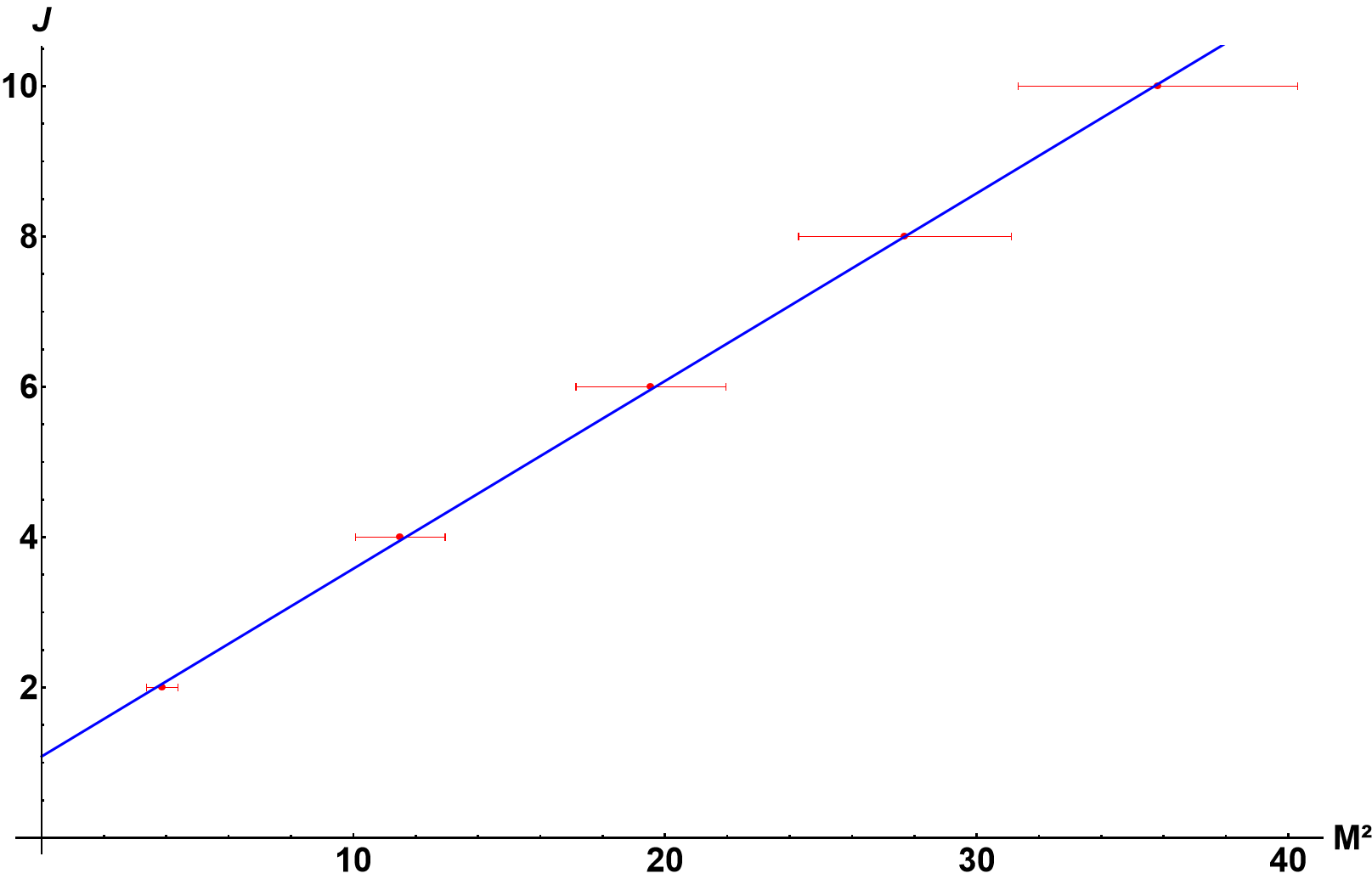}
\caption{Plot of $J\times M^2$ with $M$ expressed in GeV and $J=2$ to $10$, using the approximate anomalous dimension Eq. \eqref{anomalousdim4}, truncated at  $N=1$  with $a=2.86$, $b=1.00$. The dots with error bars represent the glueball masses shown in Table \ref{tableapprox} for $N=1$, and the straight line corresponds to a  linear fit given by $ J =  1.08\pm0.04 + (0.250\pm0.002) M^2 $ with $\chi^2/$ndf$=0.312/3=0.104$.}
\label{image4}
\end{figure}

Here, we start with the identity
\begin{equation}
    \ln x = 2 \arcsinh \left[\frac 12\left( \sqrt{x} - \frac{1}{\sqrt{x}}\right)\right]\,, 
\end{equation}
which can be expanded for small $x$ as 
\begin{eqnarray}
    \label{sqrtx-1/sqrtx1}
    \ln x &=&   \sqrt{x} - \frac{1}{\sqrt{x}}
    - \frac{1}{24}\left(  \sqrt{x} - \frac{1}{\sqrt{x}}
    \right)^3 
    +  \frac{3}{640}\left(  \sqrt{x} - \frac{1}{\sqrt{x}}
    \right)^5
    + \cdots\,\cr \cr 
    &=& 
    \sum_{k=0}^\infty \frac{(-1)^k}{2^{2k}(2k+1)k!}  
    \left(\frac 12 \right)_k 
    \left(  \sqrt{x} - \frac{1}{\sqrt{x}}
    \right)^{2k+1}\,, 
\end{eqnarray} 
where $\left(\frac 12 \right)_k$ are the Pochhammer symbols of order $k$ with argument 1/2, which first values are $\left(\frac 12 \right)_0=1$, $\left(\frac 12 \right)_1=\frac 12$, $\left(\frac 12 \right)_2=\frac 34$, $\left(\frac 12 \right)_3=\frac {15}8$, \dots. Using this expression one can rewrite the effective dimension \eqref{deltalogab} as
  \begin{eqnarray}
        \Delta  
        &=& S + a \sum_{k=0}^\infty 
        \frac{(-1)^k}{2^{2k}(2k+1)k!} 
        \left(\frac 12 \right)_k
     \left(  \sqrt{S} - \frac{1}{\sqrt{S}}
    \right)^{2k+1} 
        +b\,.  
    \end{eqnarray}
Now, truncating the series at some finite value of $k=0, 1, 2, 3, \dots$, which means truncating at odd powers $2k+1$ of the difference $\sqrt{S}-1/\sqrt{S}$, we obtain approximate expressions for $\Delta$ from which we can fit the soft pomeron trajectory and compare the obtained glueball masses with those from lattice calculations. Explicitly, we define this truncated effective dimension of high spin operators as 
  \begin{eqnarray}\label{anomalousdim4}
        \Delta_{N}
        &=&
        S + a \sum_{k=0}^N 
        \frac{(-1)^k}{2^{2k}(2k+1)k!} 
        \left(\frac 12 \right)_k 
    \left(  \sqrt{S} - \frac{1}{\sqrt{S}}
    \right)^{2k+1}  
        +b\,.
    \end{eqnarray}
Then, this anomalous truncated series HW model is characterized by the conformal dimension 
\begin{eqnarray}\label{AHWN}
\Delta_{{\rm ATSHW}}= \left\{
\begin{array}{lccccc}
4+S; &&&&& 0 \le S \le S_0; \\ 
\Delta_{N}; &&&&&  S > S_0,
\end{array}
\right. 
\end{eqnarray}
with $\Delta_{N}$ given by Eq. \eqref{anomalousdim4}. Here, we choose $S_0=0$ and $0\le S \le 10$, such that the conformal dimension for the $0^{++}$ state is $\Delta=4$ and Eq. \eqref{anomalousdim4} give the effective dimension for the states $2^{++}$ up to $10^{++}$ truncated at $N$ with $N=0,1,2,3$. Within this model, we successfully match the soft pomeron trajectory, $ J =  1.08 + 0.25 M^2 $,  in the four cases $N=0,1,2,3$. Details of the fits are presented in Table \ref{tableapprox}, together with the corresponding glueball masses and errors, the associated deviations with respect to average lattice data and the values of the anomalous dimensions for each state. Analysing these results, we see that the cases $N=0$ and $N=2$ present the smaller relative deviations with respect to average lattice output. We understand this behavior since the truncations at $N=0$ and $N=2$ contribute positively to the anomalous dimension \eqref{anomalousdim4} in contrast to the cases $N=1$ and $N=3$, once they come from an alternate series. In particular, the plots  corresponding to the masses found from the approximate anomalous effective dimension \eqref{anomalousdim4}, truncated at $N=0$ and $N=1$ are presented in Figures~\ref{image3} and \ref{image4}, respectively, with the corresponding pomeron trajectories.

\begin{table*}
\caption{\label{tableapprox}Masses of the $J^{++}$ glueball operators in GeV from $J=0$ to $J=10$,  with errors calculated according to Eq. \eqref{uncert}, from the anomalous HW  model Eq. \eqref{AHWN} considering the approximate conformal dimension  Eq. \eqref{anomalousdim4}, truncating the series at $N=0, 1, 2, 3$, with the corresponding deviations with respect to lattice data, and the anomalous  dimensions $\Delta_{\rm anom.}^N\equiv\Delta_N -(4+S)$ of the states in this model for each value of $N$. The mass of the $0^{++}$ is inserted as an input from lattice.}
\vskip0.5cm 
\tiny 
\begin{ruledtabular}
\begin{tabular}{||c||ccc|ccc|ccc|ccc||}
& & $N=0$ & & & $N=1$ & & & $N=2$ & & & $N=3$ & \\  \hline\hline
 $\alpha_0$ & &   $1.08\pm 0.28$ & & & $1.08\pm 0.04$ & & & $1.08\pm 0.34$ & & & $1.08\pm 0.29$ & 
 \\  \hline 
 $\alpha'$ & 
 & $0.25 \pm 0.01$ &&&
 $0.250 \pm 0.002$ &&& 
 $0.25 \pm 0.02$ &&&
  $0.25 \pm 0.01$ &
  \\  \hline 
 $\chi^2$/3 & 
 & 1.74 & &&
 0.104 &&&
 2.19 &&&
 1.07 &
  \\  \hline   
  $a$ && 1.38 $\pm0.26$ && & 
  2.86 $\pm0.52$ && & 
  1.40 $\pm0.26$ && & 
  3.49 $\pm0.61$ & 
 \\ \hline   
  $b$ && 2.67 $\pm0.04$ && & 
  1.00 $\pm0.26$ && & 
  2.82 $\pm0.06$ && & 
  0.10 $\pm0.38$ & 
 \\ \hline \hline 
$J^{PC}$ & $M$ & $\delta_N$ & $\Delta_{\rm anom.}^N$ & $M$ & $\delta_N$ & $\Delta_{\rm anom.}^N$  & $M$ & $\delta_N$ & $\Delta_{\rm anom.}^N$ 
& $M$ & $\delta_N$ & $\Delta_{\rm anom.}^N$ 
\\  \hline\hline 
 $0^{++}$ & $1.59\pm0.07$ & $0\%$ &  $0.0$ &  $1.59\pm0.07$  & $0\%$ & $0.0$  & $1.59\pm0.07$ & $0\%$ &  $0.0$  & $1.59\pm0.07$ & $0\%$  &  $0.0$    \\
 $2^{++}$&$2.22\pm0.13$& $6.7\%$ &  $-0.36$  & $1.97\pm0.09$   & $17\%$ &   $-1.03$ & $2.27\pm0.13$ &$4.6\%$ & $-0.21$  & $1.80\pm0.08$ & $24\%$ &     $-1.47$   \\
 $4^{++}$ & $3.34\pm0.21$ &  $9.0\%$ &  $0.74$  &  $3.39\pm0.22$  & $7.6\%$  &  $0.87$  & $3.35\pm0.21$ & $8.7\%$ &   $0.77$ & $3.41\pm0.23$ &  $7.1\%$ &  $0.94$    \\
 $6^{++}$&$4.30\pm0.27$ & $1.4\%$ &  $1.49$  & $4.42\pm0.29$  & $1.4\%$ &  $0.81$  & $4.28\pm0.27$  &  $1.8\%$ &   $1.41$  & $4.56\pm0.31$ & $4.6\%$ & $2.22$   \\
 $8^{++}$& $5.20\pm0.33$ & - & $2.08$ &   $5.26\pm0.33$    & - &  $2.25$ &  $5.17\pm0.32$ & - &   $2.00$
  & $5.40\pm0.36$ & - &   $2.69$   \\
 $10^{++}$& $6.05\pm0.38$ & - &  $2.60$  &  $5.98\pm0.36$    & - &  $2.37$  &  $6.08\pm0.38$ & - &  $2.68$
  & $5.87\pm0.34$ & -   &   $2.07$   
\end{tabular}
\end{ruledtabular}
\end{table*}

\subsection{Square root anomalous dimension}
\label{subsecsqrt}

Here, we consider another phenomenological anomalous HW model inspired by Eq. \eqref{anomalousdim4} for the anomalous dimension of the glueball operator with high spins. Considering the truncation of this equation at $N=0$, and a further approximation we write the effective dimension as 
\begin{eqnarray}\label{AHWSQRT}
\Delta_{\rm AHWSQRT}= \left\{
\begin{array}{lccccc}
4+S; &&&&& 0 \le S \le S_0; \\ 
\Delta_{\rm SQRT}; &&&&&  S > S_0,
\end{array}
\right. 
\end{eqnarray}
where 
 \begin{equation}\label{DeltaSqrt1}
        \Delta_{\rm SQRT}=S+ a\sqrt{S} +b, 
    \end{equation}
which represent square root anomalous dimensions. Here we choose $S_0=0$ and $0\le S \le 10$. The above approximation for $\Delta_{\rm SQRT}$ can be justified once we are applying this formula for spins $S$ from 2 to 10 and the correction from the inverse square root is small in this range. Using this expression for the effective dimension of the glueball operators starting with $S=2$, and minimizing the resulting linear trajectory for the corresponding glueball masses, we find the coefficients $a=1.50\pm0.27$, $b=2.0\pm0.12$. The obtained glueball masses with errors are listed in Table  \ref{tableapprox2}, together with the relative deviations with respect to average lattice data and the associated anomalous dimensions for each $J^{++}$ state. The linear coefficient of the resulting Regge trajectory, $ J =  0.90 \pm 0.30 + (0.25\pm0.01) M^2 $ with $\chi^2/$ndf$=4.24/3=1.42$, is poorer than the corresponding ones from Eq. \eqref{anomalousdim4} shown in Table \ref{tableapprox} which match the soft pomeron trajectory, $ J =  1.08 + 0.25 M^2 $. However it is remarkable that this simple model defined by Eq. \eqref{AHWSQRT} produces masses  that have smaller relative deviations with respect to average lattice data in comparison with the results of Tables \ref{tablelog} and \ref{tableapprox}.

\begin{table*}
\caption{\label{tableapprox2}Masses of the $J^{++}$ glueball operators in GeV from $J=0$ to $J=10$,  with errors calculated according to Eq. \eqref{uncert}, from the anomalous HW model, Eq. \eqref{AHWSQRT}, with square root anomalous dimensions $\Delta_{\rm SQRT}$ from Eq. \eqref{DeltaSqrt1}, together with the corresponding errors and relative deviations $\delta_{\rm SQRT}$ with respect to average lattice data, and the anomalous dimensions $\Delta_{\rm anom.}^{\rm SQRT}\equiv\Delta_{\rm SQRT} -(4+S)$ of the states $J^{++}$ in this case. The orders $\nu$ of the corresponding Bessel functions are shown for each glueball and the mass of the $0^{++}$ with uncertainties are inserted as  inputs from lattice.}
 \vskip0.5cm 
\begin{ruledtabular}
\begin{tabular}{||c||cccc||}
$J^{PC}$ & $\nu$ &  $M_{\rm SQRT}$    &  $\delta_{\rm SQRT}$  & 
$\Delta_{\rm anom.}^{\rm SQRT}$ \\  \hline
 $0^{++}$ & 2 &  $1.59\pm0.07$  &  $0\%$ & $0.0$\\
 $2^{++}$ & 4.12 & $2.40\pm0.14$  & $0.8\%$  & $0.12$\\
 $4^{++}$  & 7.00 & $3.44\pm0.21$  &  $6.3\%$ & $1.00$\\
 $6^{++}$  &  9.67  &$4.38\pm0.27$  &   $0.5\%$ &$1.67$ \\
 $8^{++}$  &  12.24  & $5.27\pm0.32$ & -  & $2.24$\\
 $10^{++}$  &  14.74   &  $6.12\pm0.37$  &  - & $2.74$
\end{tabular}
\end{ruledtabular}
\end{table*}



\section{Anomalous Linear HW  Model}
\label{secLHW}

\begin{figure}[ht]
\vskip0.5cm 
\centering
\includegraphics[width=14cm]{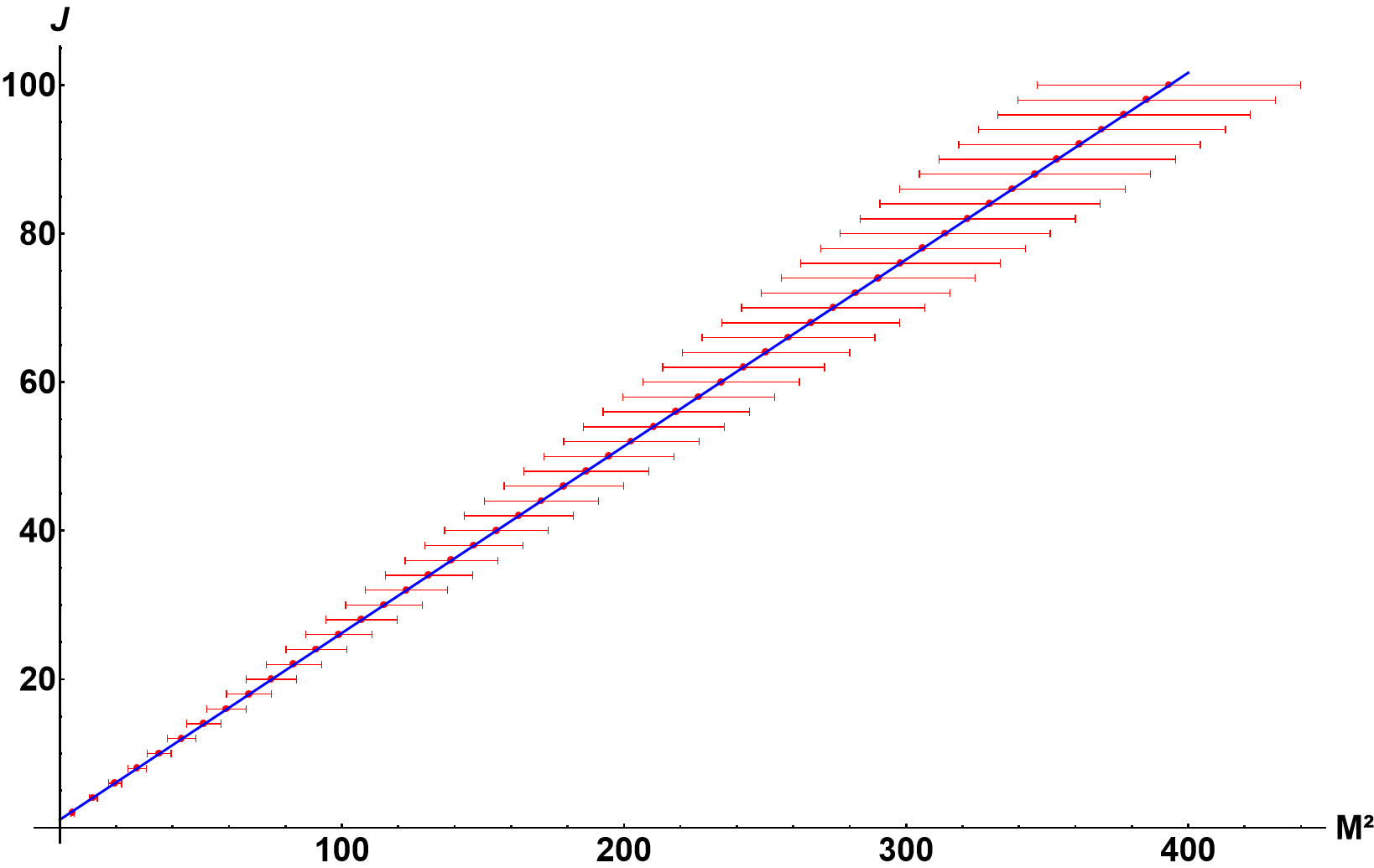}
\caption{Regge trajectory $J\times M^2$ for the anomalous linear HW model defined by Eq. \eqref{ALHW}, with $S_0=0$ and $0\le S \le 100$. The dots with error bars represent the glueball masses obtained from this equation and their values corresponding to states from $J=2$ to $J=10$ are shown in Table \ref{tablelinear}. The straight line is the corresponding linear fit in the range from $J=2$ to $J=100$, matching the soft pomeron trajectory with a very precise angular coefficient, $ J =  1.08\pm0.01 + (0.25144\pm0.00006) M^2 $ with $\chi^2/$ndf$=2.58/48=0.054$.}
\label{image5}
\end{figure}

In this section we consider another anomalous HW model inspired by the the results of the previous sections. Our goal here is to obtain an asymptotic linear Regge trajectory from such a model. In this regard, we note that the linear dependence on the spin $S$ in the effective dimension $\Delta=S + \frac{\sqrt{\lambda}}{\pi}\ln \left(\frac{S}{\sqrt{\lambda}} \right) +{\cal{O}}(S^0)$, Eq \eqref{DeltaS}, leads to parabolic Regge trajectories characteristic of the usual HW model. Then, the natural guess is to remove this linear dependence from a phenomenological effective dimension and try something of the form 
$\Delta=a\ln \left({S} \right) +b$, but this does not lead to linear Regge trajectories too. The solution is to consider an effective dimension as 
    \begin{equation}\label{newdelta}
        \Delta_{\sqrt{S}}=a \sqrt{S}+b. 
    \end{equation}
    So, the anomalous linear HW model is defined by 
\begin{eqnarray}\label{ALHW}
\Delta_{\rm ALHW}= \left\{
\begin{array}{lccccc}
4+S; &&&&& 0 \le S \le S_0; \\ 
\Delta_{\sqrt{S}}; &&&&&  S > S_0,
\end{array}
\right. 
\end{eqnarray}
where $\Delta_{\sqrt{S}}$ is given by Eq. \eqref{newdelta}, and the constants $S_0$, $a$ and $b$ will be fixed by best fit to experimental and lattice data. In order to check the possible linearity of this proposal we need to go to very high spin values. We choose $S_0=0$ and $0\le S \le 100$. The best fit for the soft pomeron trajectory from $S=2$ to $100$ with this expression is obtained for $a=6.20\pm0.26$, and $b=-3.35\pm0.11$, such that the effective dimension reads 
    \begin{equation}\label{anomdim}
        \Delta_{\sqrt{S}}=(6.20\pm0.26) \sqrt{S}-3.35\pm0.11. 
    \end{equation}
The resulting trajectory is plotted in Figure \ref{image5}, and is in perfect agreement with the experimental soft pomeron trajectory, $ J =  1.08\pm0.01 + (0.25144\pm0.00006) M^2 $ with $\chi^2/$ndf$=2.58/48=0.054$. The masses for glueballs obtained from this effective dimension are shown in Table \ref{tablelinear}, together with the relative deviations with respect to average lattice data, and the corresponding anomalous dimensions of the states $J^{++}$. 

\begin{table*}
\caption{\label{tablelinear}Masses of the $J^{++}$ glueball operators in GeV from $J=0$ to $J=10$ within the anomalous linear HW  (ALHW) model, Eq. \eqref{ALHW}, with errors calculated according to Eq. \eqref{uncert}. The orders $\nu$ of the corresponding Bessel functions are shown for each glueball state and the mass of the $0^{++}$ with uncertainties are inserted as  inputs from lattice.  We also show the relative deviations $\delta_{\rm ALHW}$ compared with average lattice data, and the anomalous dimensions $\Delta_{\rm anom.}^{\rm ALHW}\equiv\Delta_{\rm anom.} -(4+S)$ of the glueball operators $J^{++}$ in this model.}
\vskip0.5cm 
\begin{ruledtabular}
\begin{tabular}{||c||cccc||}
$J^{PC}$ & $\nu$ & $M_{\rm ALHW}$ & $\delta_{\rm ALHW}$ & $\Delta_{\rm anom.}^{\rm ALHW}$  
\\ \hline
 $0^{++}$ & 2 
 &$1.59\pm0.07$  &$0\%$ &$0.0$
 \\
 $2^{++}$ & 3.42 & $2.14\pm0.13$ &$10.1\%$ &$-0.58$
 \\
 $4^{++}$ & 7.05  & $3.46\pm0.21$ &$5.72\%$ &  $1.05$
 \\
 $6^{++}$ & 9.84  & $4.44\pm0.27$& $1.83\%$ & $1.84$
 \\
 $8^{++}$ & 12.19  & $5.96\pm0.31$ & - &$2.19$
 \\
 $10^{++}$ & 14.26  & $6.59\pm0.35$ & - & $2.26$
\end{tabular}
\end{ruledtabular}
\end{table*}

Comparing the above results with the ones from Eq. \eqref{anomalousdim4}, one may wonder whether a contribution of the inverse of the square root of $S$ would spoil the linearity just found. Actually, considering a model with effective dimension given by $\Delta_{\tilde N=0}=a(\sqrt{S}-{1}/{\sqrt{S}})+b$ without the linear term, it is  possible to obtain an asymptotic linear Regge trajectory in this case. Fitting the states from  $J=2$ to $J=10$ with coefficients  $a=5.40\pm0.26$ and $b=0.86\pm0.04$, we also match exactly the soft pomeron Regge trajectory in very good approximation, $ J =  1.08\pm0.05 + (0.252\pm0.002) M^2 $ with $\chi^2/$ndf$=0.260/3=0.087$. The masses obtained for these states are $\{1.85\pm0.12, 3.42\pm0.21, 4.44\pm0.28, 5.25\pm0.33, 5.93\pm0.37 \}$ GeV, with relative deviations $\{22.3, 6.81, 1.83\}\%$ with respect to average lattice data. This suggests that the results obtained from Eq. \eqref{anomdim} are better. The other powers $N=1,2,3$ of the difference $\sqrt{S}-{1}/{\sqrt{S}}$ discussed in the previous section, do not lead to asymptotic linear trajectories when high spins are considered, since they  grow with $S^{(2N+1)/2}$. 


\section{Summary and Conclusions}
\label{conclusions}

\subsection{AHW models}

In this work we propose anomalous HW  models in four different formulations where we modify the conformal dimension for high spins $S$ operators by  introducing anomalous dimensions inspired by a semi-classical limit of gauge/string dualities. For low spins the dimension of the spin operators are conformal $(\Delta=S+4)$ as in the original HW model. For high spins, these operators acquire anomalous dimensions in the form of a logarithm of $S$. Disregarding a complicated intermediate non-perturbative spin dependence, we take these results as a prescription to build up anomalous HW models. In general, to compare these models with experimental and lattice data, we adopt the strategy of first fitting the pomeron trajectory predicting glueball masses to be compared with lattice data. The effective dimensions of these four formulations of the anomalous HW models are:
\begin{itemize}
    \item Logarithm AHW 
\begin{eqnarray}\label{AHWlog}
\Delta_{\rm AHWlog}= \left\{
\begin{array}{lccccc}
4+S; &&&&& 0 \le S \le S_0; \\ 
S + a \ln \left({S} \right) + b; &&&&&  S > S_0;
\end{array}
\right. 
\end{eqnarray}
\item Truncated Series AHW  
\begin{eqnarray}\label{ATSHW}
\Delta_{\rm ATSHW}= \left\{
\begin{array}{lccccc}
4+S; &&&&& 0 \le S \le S_0; \\ 
 \Delta_N; &&&&&  S > S_0; 
\end{array}
\right. 
\end{eqnarray}
where 
\begin{equation}
    \Delta_N=S + a \sum_{k=0}^N 
        \frac{(-1)^k}{2^{2k}(2k+1)k!} 
        \left(\frac 12 \right)_k 
    \left(  \sqrt{S} - \frac{1}{\sqrt{S}}
    \right)^{2k+1}  
        +b; 
\end{equation}
\item Square root AHW 
\begin{eqnarray}\label{AHWSQRTR}
\Delta_{\rm AHWSQRT}= \left\{
\begin{array}{lccccc}
4+S; &&&&& 0 \le S \le S_0; \\ 
 S+ a\sqrt{S} +b\,; &&&&&  S > S_0\; 
\end{array}
\right. 
\end{eqnarray}
\item Linear AHW 
\begin{eqnarray}\label{LAHW}
\Delta_{\rm ALHW}= \left\{
\begin{array}{lccccc}
4+S; &&&&& 0 \le S \le S_0; \\ 
 a \sqrt{S}+b\,; &&&&&  S > S_0. 
\end{array}
\right. 
\end{eqnarray}
\end{itemize}

\subsection{Log models}
\label{logmodels}

In Section \ref{subseclog}, we consider the specific case of logarithm anomalous dimensions, Eq. \eqref{AHWlog}, where we study some numerical fits to reobtain the soft pomeron trajectory and even glueball masses for the states $J=2$ to $J=10$. In particular, in this section we analyse four different fits with the following phenomenological anomalous dimensions with the corresponding spins interval for which these expressions apply:   
\begin{eqnarray}
\Delta_{\rm log1}= 
S +(1.92\pm0.36)\ln{(S)}+2.13\pm0.07;  \qquad\qquad (2\le S \le 10);
\end{eqnarray}
\begin{equation}
\Delta_{\rm log2}= S +(2.32\pm0.25)\ln{(S)}+1.28\pm0.21; \qquad\qquad (4\le S \le 10);
\end{equation}
\begin{equation}
\Delta_{\rm log3}=S+(2.34\pm0.04)\ln(S)+0.51\pm0.02; \qquad\qquad (6\le S \le 10);
\end{equation}
\begin{equation}\label{deltalog4}
\Delta_{\rm log4}= S +(2.3\pm0.8)\ln{(S)}+2.5\pm1.1; \qquad\qquad (2\le S \le 10). 
\end{equation}
In the first three log anomalous dimensions, we follow the procedure of fitting the pomeron trajectory and then compare the glueball mass outputs with lattice data. In the fourth case we reverse this approach and fit the lattice masses and then compare the resulting Regge trajectory with that of the pomeron. With the first two fits we could reproduce precisely the pomeron trajectory $J=1.08+0.25M^2$, while the third we found $J=1.10+0.27M^2$. Actually, in the first fit we found better results regarding the deviations from the pomeron trajectory and smaller $\chi^2$/ndf. In the fourth log fit we found good masses as expected but a poorer Regge trajectory $J = 0.93  + 0.22 M^2$ when compared with the pomeron. The first two log fits also present better results for the pomeron trajectory and glueball masses than the original HW and SW models. 

\subsection{Truncated series models}

In Section \ref{subsecapprox} we discuss approximations to the logarithm anomalous dimensions in the form of an infinite series of odd powers of $(\sqrt{S}-1/\sqrt{S})^{2N+1}$ which are truncated at $N=0, 1, 2, 3$, as shown in Eq. \eqref{ATSHW}. In these four cases they all fit the pomeron trajectory $J=1.08+0.25M^2$ with different precisions. Regarding the glueball masses preditions when compared with lattice data, the cases with $N=0$ and $N=2$  present smaller relative deviations than the cases $N=1$ and $N=3$. In particular, for $N=0, 1, 2$ the total relative deviations are smaller than that of the original HW model, while for $N=3$ they are of the same order of magnitude. These results are also better than that of the original SW model. 

\subsection{Square Root Anomalous dimensions}

Inspired by the results of the truncated series AHW models, in Section \ref{subsecsqrt} we discuss an AHW model with square root anomalous dimensions, Eq. \eqref{AHWSQRTR}. This model gives as output the Regge trajectory $J=0.90 + 0.25M^2$, which does not fit exactly the soft pomeron. On the other hand, the predicted  glueball masses by this model presents smaller deviations with respect to lattice data than the models with truncated series and the logarithm models AHW$_{\rm log1}$, AHW$_{\rm log2}$, and AHW$_{\rm log3}$. Naturally, the glueball masses predicted by this model are much better than the ones from the original HW and SW models when compared with lattice data. 

\subsection{Anomalous Linear HW Model}

In Section \ref{secLHW}, we propose an anomalous linear HW  model, Eq. \eqref{LAHW}, leading to asymptotic linear Regge trajectories even for very high spins ($J=100$). In contrast, the original HW model is well known for producing non-linear Regge trajectories. The main modification introduced in the linear model is that we take the dimension of the glueball operators $\Delta=a\sqrt{S}+b$ without the linear term $S$ present in all anomalous models discussed above and also in the original HW model. The obtained Regge trajectory in the linear AHW model fits the soft pomeron  with very high precision $ J =  1.08\pm0.01 + (0.25144\pm0.00006) M^2 $ with $\chi^2/$ndf$=2.58/48=0.054$, and the glueball masses compare well to lattice data with total deviation of the order of some above anomalous HW models, better than the original HW and SW models. 

\subsection{Conclusions}

In conclusion, the anomalous HW models presented here show a significant improvement with respect to the original HW model in general and with respect to the original SW model regarding glueball spectra and the pomeron trajectory. This procedure of modifying the HW model with the inclusion of anomalous dimensions might be useful for other hadrons as mesons and baryons. This is presently under investigation.

\section*{Acknowledgments}
We would like to thank an anonymous referee for the criticisms and suggestions raised which helped us improve the quality of this work, especially in its numerical analysis. 
R.A.C.-S. is supported by  Conselho Nacional de Desenvolvimento Científico e Tecnológico (CNPq) and Coordenação de Aperfeiçoamento de Pessoal de Nível Superior (CAPES) under finance code 001. H.B.-F. is partially supported by Conselho Nacional de Desenvolvimento Cient\'{\i}fico e Tecnol\'{o}gico (CNPq) under grants $\#$s 311079/2019-9 and 310346/2023-1.



\end{document}